\documentclass[11pt]{article}
\usepackage{amsfonts}

\usepackage[final]{acl}
\usepackage{hyperref}
\usepackage{xcolor}
\usepackage{times}
\usepackage{latexsym}
\usepackage{amsmath}
\usepackage{multirow}
\usepackage[T1]{fontenc}
\usepackage{multirow}
\usepackage{tabularx}
\usepackage[utf8]{inputenc}

\usepackage{microtype}
\usepackage[table]{xcolor}

\usepackage{inconsolata}

\usepackage{graphicx}
\usepackage{booktabs}

\title{Foley-Omni: A Unified Multimodal Generation Model from Task-Level Audio Synthesis to Complete Video Soundtrack Generation}

\author{
  \textbf{Ye Tao\textsuperscript{1,2,3}},
  \textbf{Lupeng Liu\textsuperscript{2,4}},
  \textbf{Xuenan Xu\textsuperscript{5}},
  \textbf{Jiasun Feng\textsuperscript{2}},
  \textbf{Jiarui Wang\textsuperscript{2}},
\\
  \textbf{Ying Qin\textsuperscript{4}},
  \textbf{Shuiyang Mao\textsuperscript{2}},
  \textbf{Wei Liu\textsuperscript{2}},
  \textbf{Shuai Wang\textsuperscript{1,*}}
\\
\\
  \textsuperscript{1}School of Intelligence Science and Technology, Nanjing University,
  \textsuperscript{2}Video Rebirth,
\\
   \textsuperscript{3}Shanghai Jiao Tong University
  \textsuperscript{4}Beijing Jiaotong University,
  \textsuperscript{5}Shanghai AI Laboratory
\\
 \texttt{taoye0402@gmail.com}
}

\begin{document}
\maketitle

\begin{abstract}
Recent unified audio generation models can support diverse tasks across speech, sound effects, and music, but most of them still focus on isolated task-level synthesis. However, real video production often requires multiple components of a complete audio track to be generated jointly and consistently for the same video.
We present Foley-Omni, a unified multimodal audio generation model that extends isolated task-level synthesis to complete video soundtrack generation by jointly modeling speech, sound effects, and music within a shared latent generation process.
To support training and reproducible evaluation, we develop an audiovisual data curation pipeline and introduce V2ST-Bench, a benchmark for holistic video soundtrack generation evaluation.
Experiments show that Foley-Omni achieves competitive performance with expert systems on individual synthesis tasks, while improving speech intelligibility, audiovisual consistency and perceptual quality for mixed soundtrack generation. The project page can be accessed at \href{https://ty0402.github.io/Foley-omni-Web/}{\textcolor{blue!60!black}{Project Page}}.
\end{abstract}

\section{Introduction}


Modern video creation requires more than visually realistic content. A complete video should be accompanied by a coherent audio track, where speech is intelligible and synchronized with speakers, sound effects match visible events, and music supports the rhythm and atmosphere of the scene. Recent closed-source systems, such as Google's Veo3~\cite{veo3_techreport}, have moved toward unified audiovisual generation, enabling joint synthesis of videos with speech, sound effects, and music.
In contrast, academic audio generation research is still largely organized around individual synthesis tasks, including text-to-audio (TTA), text-to-speech (TTS), text-to-music (TTM), video-to-audio (V2A) and visual text-to-speech (VisualTTS).

\begin{figure}[!t]
  \centering
  \includegraphics[width=1.0\columnwidth]{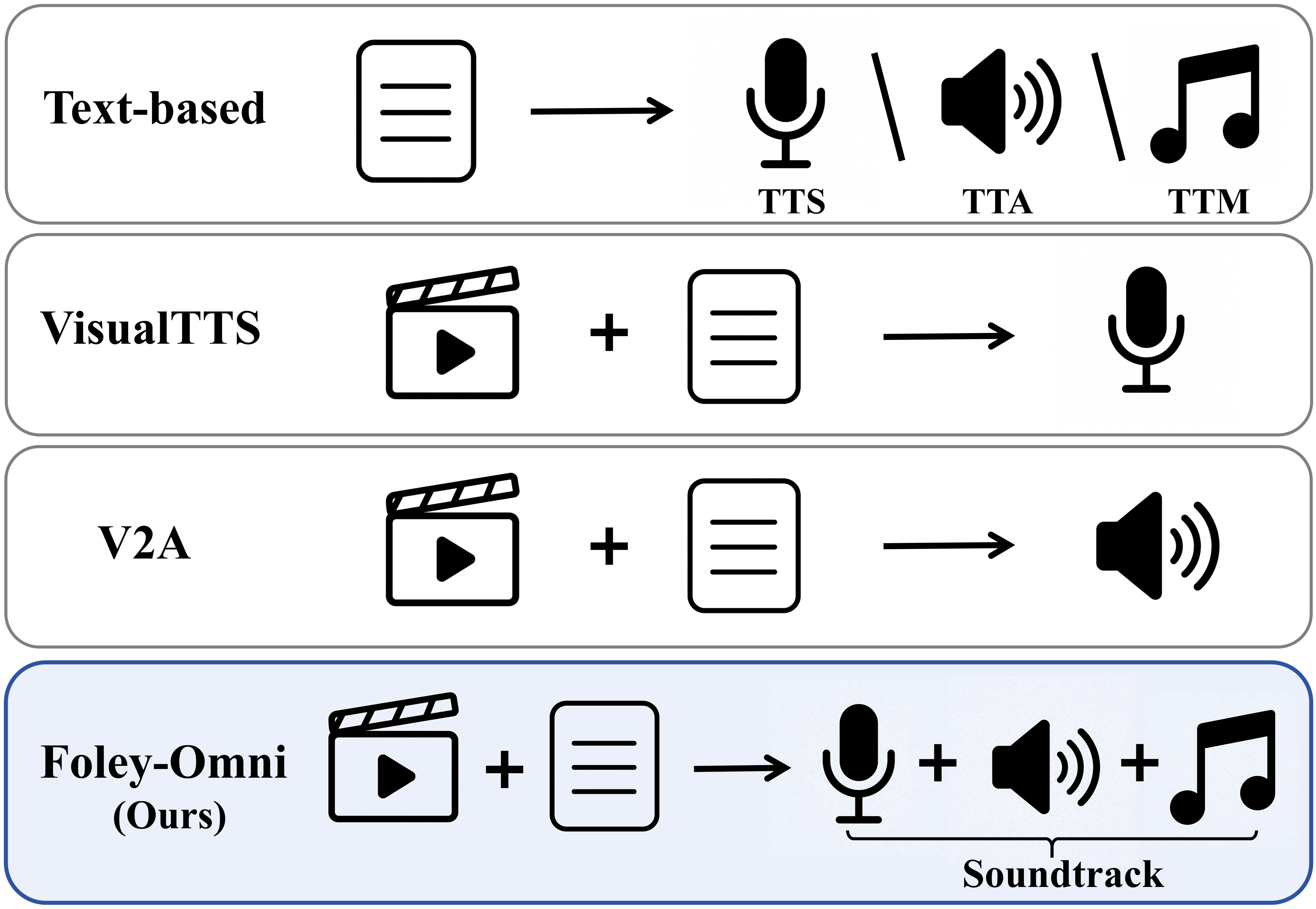}
  \vspace{-1.5em}
  \caption{Overview of Foley-Omni. Foley-Omni supports task-level audio synthesis and further generates mixed audio for videos within a unified framework.}
  \label{fig:intro}
  \vspace{-1.0em}
\end{figure}

Recent unified audio generation models, such as AudioX~\cite{tian2025audiox}, show that a single model can support multiple audio domains and task formulations.
However, their unification is still mostly demonstrated at the task level: the model can support multiple tasks, but each generation process typically focuses on a single audio domain.
This task-level unification does not fully capture the requirements of realistic \textbf{Video-to-Soundtrack (V2ST)} generation, where speech, sound effects, and music need to be jointly generated while remaining temporally and semantically consistent with the input video.

This challenge of domain isolation similarly plagues current video-conditioned models.
V2A models emphasize non-speech sounds aligned with visual events, whereas VisualTTS and dubbing models emphasize speech intelligibility and lip synchronization.
Recent work, such as DualDub~\cite{tian2025dualdub} and VSSFlow~\cite{cheng2025vssflow}, aims to unify video-conditioned sound and speech generation.
However, these systems are not open-sourced, and their evaluations still mainly rely on single-task benchmarks. 

Another critical bottleneck for the V2ST task lies in data and evaluation.
Large-scale audiovisual datasets are often weakly labeled or temporally misaligned.  
More importantly, they typically lack structured annotations for the distinct audio components that coexist within a video, which limits their applicability to complete soundtrack generation.
Meanwhile, the community lacks a publicly accessible high-quality benchmark for V2ST evaluation.
This makes it difficult to systematically evaluate V2ST performance.

To this end, we tackle these challenges from both data and model perspectives.
On the data side, we develop an audiovisual data pipeline that transforms weakly labeled audiovisual data into structured training samples with component-level annotations.
Based on the pipeline, we further propose \textbf{V2ST-Bench}, a public benchmark for systematically evaluating complete video soundtrack generation.

On the modeling side, we present \textbf{Foley-Omni}, a unified multimodal framework for general audio generation from video and text.
As shown in Figure~\ref{fig:intro}, Foley-Omni unifies task-level synthesis and complete video soundtrack generation.
More importantly, it addresses the V2ST task by jointly generating multiple audio components as a coherent track, instead of generating them separately and mixing afterward.
Experiments show that Foley-Omni remains competitive on task-level synthesis while achieving stronger intelligibility and audiovisual consistency in complete soundtrack generation.
Our contributions are summarized as follows:

\begin{itemize}
  \item We propose \textbf{Foley-Omni}, a unified multimodal audio generation model that supports speech, sound effects, and music across TTA, TTS, TTM, V2A, and VisualTTS.


  \item We introduce a curriculum learning strategy that bridges task-level synthesis and complete soundtrack generation, gradually extending the model from general audio priors to complete video soundtrack generation.

  \item We develop an audiovisual data curation pipeline and \textbf{V2ST-Bench}, establishing a structured dataset for training and reproducible evaluation for complete video soundtrack generation.

  \item Extensive experiments show that Foley-Omni achieves competitive performance across individual synthesis tasks and achieves stronger intelligibility and audiovisual consistency in mixed audio generation.
\end{itemize}

\section{Related Work}

\subsection{Unified Audio Generation}

Audio generation has developed from several task-specific directions. For TTA, AudioLDM~2~\cite{liu2024audioldm2} shows the effectiveness of latent diffusion for general audio synthesis, while Tango~2~\cite{majumder2024tango} improves text-audio alignment via preference optimization. For TTS, CosyVoice~\cite{du2025cosyvoice} and F5-TTS~\cite{chen2025f5} achieve strong speech synthesis through language-modeling objectives and conditional flow matching. TTM models~\cite{copet2023simple} further extend text-guided generation to musical structure, rhythm, and style. Despite their strong task-specific performance, these systems are developed under separate formulations and do not support unified modeling across speech, sound effects, and music.

Recent unified models aim to reduce this fragmentation by covering multiple audio domains within a single framework. AudioX~\cite{tian2025audiox} supports diverse audio generation tasks across multiple input modalities, while UniFlow-Audio~\cite{xu2025uniflow} studies a unified flow-matching formulation for speech, music, and sound effects. Audio-Omni~\cite{tian2026audio} further integrates audio understanding, editing, and generation within one framework. In parallel, WavJourney~\cite{liu2025wavjourney} explores composite audio generation by using an LLM to plan audio scripts and coordinate multiple expert modules. 
However, they mainly focus on task-level synthesis or module-level composition, and cannot end-to-end generate a complete soundtrack for a given video.

\subsection{Video-Conditioned Audio and Speech Generation}

\begin{figure*}[t]
  \centering
  \includegraphics[width=\textwidth]{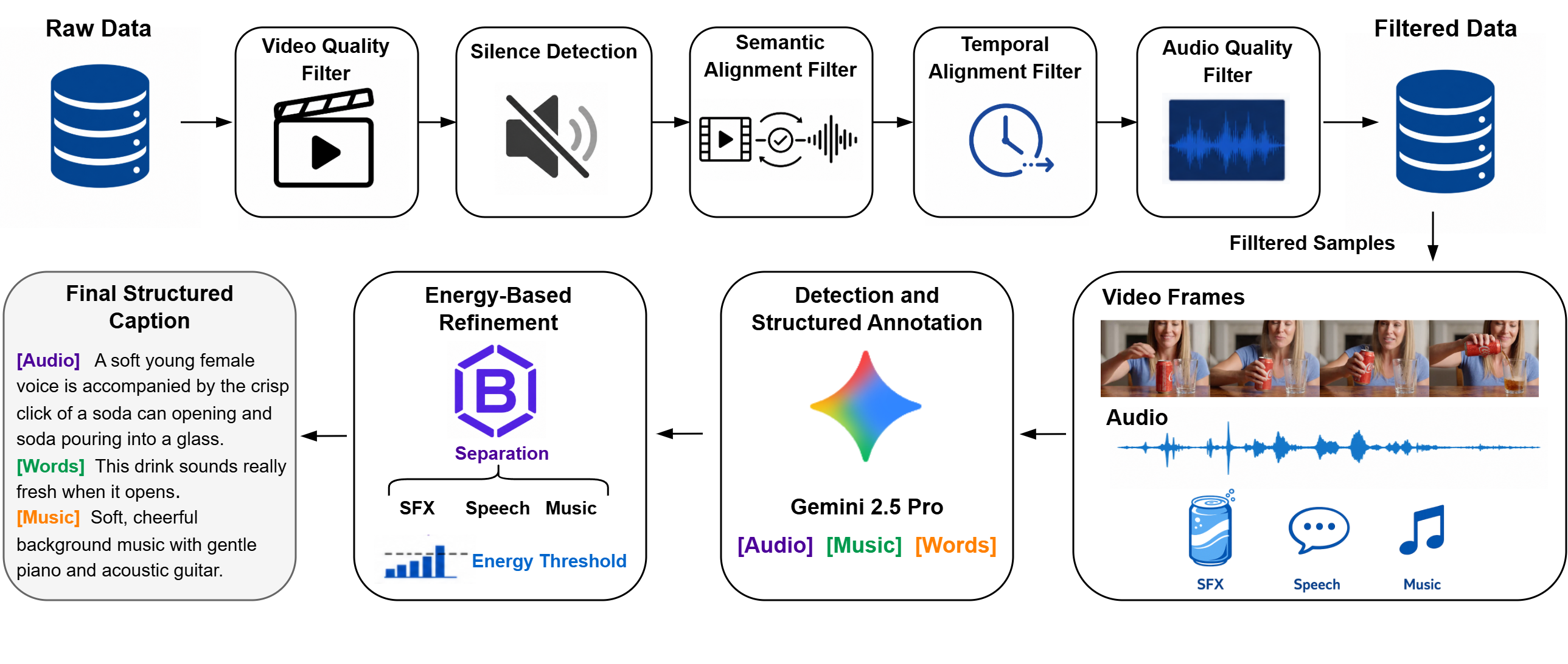}
  \caption{Audiovisual data curation pipeline. The pipeline combines quality filtering, Gemini-based structured labeling, and Bandit component verification to produce unified video-audio-text tuples for training and evaluation.}
\vspace{-0.5 em}
  \label{fig:data-pipeline}
\end{figure*}

Video-grounded audio generation mainly follows two directions. VisualTTS and video dubbing methods use visual cues to guide speech generation, with works such as V2C-Net~\cite{chen2022v2c} exploring visual voice cloning and EmoDubber~\cite{cong2025emodubber} improving expressive dubbing through disentangled modeling of speaker style, emotion, and linguistic content. However, they are strictly limited to clean speech tracks and neglect sound effects or music. In contrast, V2A methods synthesize sounds aligned with visual events; FoleyCrafter~\cite{zhang2026foleycrafter} adapts pretrained audio generation models for video-aligned Foley synthesis, while MMAudio~\cite{cheng2025mmaudio} uses a multimodal DiT framework for audio generation from text and video. These methods achieve strong results on environmental and action-related sounds, but they usually lack linguistic control and cannot generate intelligible speech.

Recent studies have started to bridge the gap between V2A and VisualTTS. DualDub uses separate branches for dubbed speech and sound effects with fusion modules  to show the feasibility of V2ST, though it still fundamentally relies on two separate generation processes. VSSFlow~\cite{cheng2025vssflow} unifies video-conditioned sound and speech generation by investigating different condition injection strategies, while AudioGen-Omni~\cite{wang2026audiogen} employs multiple encoders and a multi-modal DiT architecture to support diverse tasks. DeepSound~\cite{liang2025deepsound} further explores reasoning-based generation to reduce conflicts among audio components.
Despite these advances, complete video soundtrack generation remains underexplored.
Moreover, these systems are not publicly available, and their evaluations still mainly rely on single-task benchmarks, hindering systematic performance comparisons across complete video soundtrack generation systems.

\section{Data Pipeline and V2ST-Bench}
\subsection{Audiovisual Data Curation Pipeline}
\label{sec:datapipeline}

Open-source audiovisual datasets, such as VGGSound~\cite{chen2020vggsound}, are not well suited for video-text-conditioned audio generation.
Their textual annotations are typically coarse-grained, with weak semantic correspondence or poor temporal synchronization between visual and audio streams, resulting in noisy supervision signals.
Training on such data may result in unreliable synchronization and degraded mixed audio generation quality.
Therefore, we build an audiovisual data curation pipeline that converts weakly labeled audio-video pairs into structured video-audio-text samples with explicit fields for speech, sound effects, and music. Figure~\ref{fig:data-pipeline} illustrates the overall pipeline.
The complete details of this pipeline, including the filtering metrics, annotation prompts and examples, are provided in Appendix~\ref{app:pipeline-details}.

We first filter low-quality audio-video clips before annotation.
The filtering stage removes clips with silence, low visual resolution, poor audio quality, weak audiovisual semantic consistency, or unreliable synchronization.
This step reduces data noise, yielding high-quality clips for video-grounded audio generation.

\begin{figure*}[t]
  \centering
  \includegraphics[width=\textwidth]{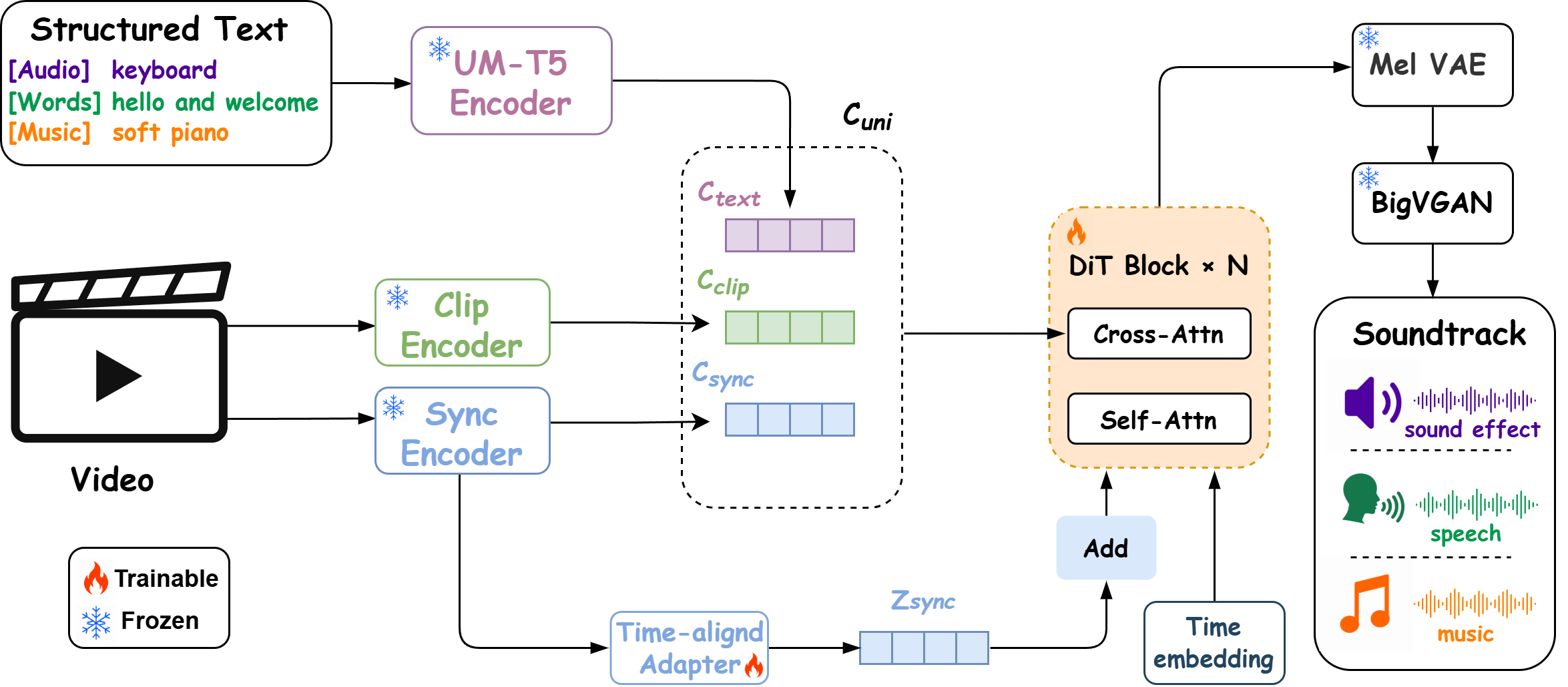}
  \caption{Overall architecture of Foley-Omni. Structured text, CLIP features, and synchronization-aware visual features form a unified multimodal context for the DiT backbone. Synchronization features are injected through both cross-attention and an extra additive path to strengthen temporal alignment.}
  \label{fig:overall-arch}
  \vspace{-0.5 em}
\end{figure*}

For the retained clips, we use Gemini 2.5 Pro~\cite{comanici2025gemini} to generate structured annotations for both audio and video.
The annotation uses explicit field tags to separate different audio components, including speech content, scene-grounded sound events, and music.
This provides detailed textual labels for the video-audio pairs.
However, these annotations may still include inaudible sound labels inferred from visual cues, a common issue in multimodal audio annotation~\cite{dai2026omni2sound}.
To reduce this visual bias, we separate the original soundtrack into speech, sound effects, and music tracks using Bandit~\cite{watcharasupat2024remastering}.
A label is removed when the energy of its corresponding separated track falls below a threshold.
After filtering, annotation, and verification of audio components, each sample is organized as a unified video-audio-text tuple $(v_i, a_i, \hat{\mathbf{S}}_i)$, where $\hat{\mathbf{S}}_i$ denotes the verified structured text.
To ensure consistent supervision across heterogeneous datasets, we apply the same pipeline to all video-based training samples, including existing datasets, before model training.
This comprehensive data-curation pipeline yields approximately 2.0M video-audio-text tuples for training.

\subsection{V2ST-Bench}

Existing benchmarks mostly evaluate single-component audio generation tasks such as V2A or VisualTTS.
Although DualBench~\cite{tian2025dualdub} moves toward the joint evaluation of speech and sound effects, its reliance on copyrighted animation sources and an internal preprocessing pipeline limits data accessibility and reproducibility.
To support reproducible evaluation for complete video soundtrack generation, we construct \textbf{V2ST-Bench}.
V2ST-Bench focuses on clips where speech and non-speech audio coexist within the same scene.
Candidate samples are drawn from the curated pool produced by our data pipeline, including open-source speaker video datasets and videos collected from publicly available web sources.

Specifically, we retain clips whose verified annotations contain at least two components among speech, music, and sound effects. 
These candidates are then manually reviewed for audiovisual consistency, annotation accuracy, and suitability for mixed soundtrack evaluation.
The final benchmark contains 300 high-quality video-audio-text tuples. 
Additional statistics are provided in Appendix~\ref{app:v2st-details}.
We will release annotations, metadata, and processing scripts to support reproducible evaluation of complete video soundtrack generation.

\section{Method}

\subsection{Overview}

Figure~\ref{fig:overall-arch} shows the overall architecture of Foley-Omni. 
Foley-Omni uses structured text to unify semantic control over speech, sound effects, and music.
In addition, CLIP~\cite{radford2021clip} and Synchformer~\cite{iashin2024synchformer} features are used to provide visual semantics and temporal cues from the video.
These conditions are injected into a diffusion Transformer (DiT) backbone, which directly generates a mixed audio track in the latent space. 
Following MMAudio~\cite{cheng2025mmaudio}, we use its frozen Mel VAE to transform the audio into the latent space, and a BigVGAN vocoder~\cite{leebigvgan} to convert the decoded mel-spectrograms into final waveforms. 
The framework supports task-level synthesis, including TTA, TTS, TTM, V2A, and VisualTTS, and further extends to complete video soundtrack generation through a curriculum learning strategy.

\subsection{Structured Multimodal Conditioning}

Foley-Omni receives heterogeneous conditions from text and video.
To enable shared text conditioning for different tasks, we use explicit field tags to organize different textual conditions into a single structured text description.
Specifically, \texttt{[WORDS]} denotes the spoken content, \texttt{[AUDIO]} describes sound events, and \texttt{[MUSIC]} specifies music elements. 
Each field can be left empty when the corresponding audio component is absent.
This design allows TTA, TTS, TTM, V2A, VisualTTS, and complete video soundtrack generation to share the same textual interface, while the explicit tags provide type information for speech, sound effects, and music.
Unlike other general models~\cite{wang2026audiogen} that use separate text encoders for different tasks, we encode the structured sequence with a shared UM-T5 encoder~\cite{chungunimax}, so transcripts, sound descriptions, and music prompts are mapped into a shared semantic space.

For video input, Foley-Omni extracts two complementary visual conditions.
CLIP features capture scene-level semantics while Synchformer features capture temporal information related to lip motion and action boundaries.
The semantic stream guides the audio content to be generated, while the synchronization stream provides timing cues to align audio events with the video.
After projection, the text representation $\mathbf{C}_{\mathrm{text}}$, CLIP features $\mathbf{C}_{\mathrm{clip}}$, and synchronization features $\mathbf{C}_{\mathrm{sync}}$ are concatenated into a unified multimodal context:
\begin{equation}
\mathbf{C}_{\mathrm{uni}}
=
[\mathbf{C}_{\mathrm{text}};
 \mathbf{C}_{\mathrm{clip}};
 \mathbf{C}_{\mathrm{sync}}].
\end{equation}

\subsection{Hybrid Condition Injection}

Cross-attention provides flexible semantic conditioning, but it does not explicitly align video frames with audio along the time axis.
To provide stronger temporal grounding, Foley-Omni converts the Synchformer features into a time-aligned representation whose sequence length matches that of the audio latents, using an adapter composed of interpolation and multi-layer projection:
\begin{equation}
\mathbf{Z}_{\mathrm{sync}}
=
\operatorname{Adapter}
(\mathbf{C}_{\mathrm{sync}}, L),
\end{equation}
where $L$ denotes the audio latent length.
The aligned synchronization representation is then added to the noisy audio latent $\mathbf{x}_{t}$ at the denoising timestep $t$:
\begin{equation}
\tilde{\mathbf{x}}_{t}
=
\mathbf{x}_{t} + \mathbf{Z}_{\mathrm{sync}}.
\end{equation}

The Transformer backbone takes $\tilde{\mathbf{x}}_{t}$, $t$ and $\mathbf{C}_{\mathrm{uni}}$ as input, where $t$ is encoded to modulate the backbone through adaptive layer normalization (AdaLN), and $\mathbf{C}_{\mathrm{uni}}$ is incorporated into the backbone via cross-attention.
Within each block, self-attention models temporal dependencies among audio latent tokens, while cross-attention injects the structured multimodal conditions.
The conditional velocity predictor can be written as:
\begin{equation}
\hat{\mathbf{v}}
=
\mathbf{v}_{\theta}
(\tilde{\mathbf{x}}_{t}, t, \mathbf{C}_{\mathrm{uni}}).
\end{equation}

This design uses two complementary condition-injection paths.
The unified context $\mathbf{C}_{\mathrm{uni}}$ is incorporated through cross-attention for flexible semantic conditioning, while the time-aligned representation $\mathbf{Z}_{\mathrm{sync}}$ is added to the audio latent sequence to provide fine-grained temporal guidance.

\subsection{Flow Matching Training Objective}

We train Foley-Omni with conditional flow matching~\cite{lipmanflow} in the audio latent space.
We define a linear interpolation path between noise $\mathbf{x}_{0}\sim\mathcal{N}(\mathbf{0},\mathbf{I})$ and data $\mathbf{x}_{1}$, which is the audio latent encoded by the VAE:
\begin{equation}
\mathbf{x}_{t}
=
(1-t)\mathbf{x}_{0}
+
t\mathbf{x}_{1},
\qquad t \in [0,1].
\end{equation}
The corresponding target velocity is:
\begin{equation}
\mathbf{v}^{*}
=
\frac{\mathrm{d}\mathbf{x}_{t}}{\mathrm{d}t}
=
\mathbf{x}_{1}
-
\mathbf{x}_{0}.
\end{equation}

Using the conditional velocity predictor introduced above, the training objective is:

\begin{equation}
\mathcal{L}
=
\mathbb{E}
\left[
\left\|
\mathbf{v}_{\theta}
(\tilde{\mathbf{x}}_{t}, t, \mathbf{C}_{\mathrm{uni}})
-
(\mathbf{x}_{1}-\mathbf{x}_{0})
\right\|_{2}^{2}
\right].
\end{equation}

Since speech, sound effects, and music are generated jointly by the same backbone, Foley-Omni models their co-occurrence and balance within a unified latent generation process, avoiding the need for separate generation and post-hoc mixing.

\subsection{Curriculum Learning Strategy}
\label{sec:progressive-training}
Training Foley-Omni requires balancing heterogeneous generation abilities across speech, sound effects, and music under text and video conditions.
In our preliminary experiments, directly mixing all data leads to task interference: in mixed audio generation, speech becomes less intelligible and sound effects are suppressed by other components.
To mitigate this issue, we train Foley-Omni progressively: the model first learns text-driven generation priors for general audio, then learns to use video conditions for audiovisual alignment, and finally adapts to complete soundtrack generation.

\paragraph{Stage 1: Text-driven audio pretraining.}
We first train Foley-Omni on TTA, TTS, and TTM data.
This stage builds general audio generation priors across speech, sound effects, and music, and allows the model to learn the structured text condition before introducing video conditions.

\paragraph{Stage 2: Video-conditioned expansion.}
We then introduce V2A and VisualTTS data to teach the model how to use visual information.
This stage extends Foley-Omni from text-driven generation to video-conditioned generation, where CLIP features of video provide scene-level semantic guidance and synchronization features provide timing cues for speech and sound-producing events.

\paragraph{Stage 3: Complete soundtrack finetuning.}
We finally finetune on mixed audiovisual samples with coexisting audio components.
This stage adapts Foley-Omni to complete soundtrack generation, where speech intelligibility, audiovisual synchronization, and mixture balance are optimized jointly.
To mitigate catastrophic forgetting of task-level abilities learned from previous stages, we retain a portion of single-task data during this stage.

\section{Experimental Setup}

\subsection{Training Settings}
\label{sec:training-settings}

Foley-Omni is trained across three sequential stages using a comprehensive multimodal corpus of approximately 2.7M pairs spanning six task groups.
Detailed data compositions and training settings for each stage are provided in Appendix~\ref{app:data-details} and~\ref{app:trainconfig}.

\subsection{Evaluation Settings}
\label{sec:baselines-metrics}

Beyond validating  task-level synthesis capabilities  (TTA, TTS, TTM, V2A, VisualTTS) on standard benchmarks, we conduct our main evaluation on V2ST-Bench to assess whether the model can generate a complete soundtrack that remains semantically consistent with the text condition, temporally synchronized with the video, and coherent as a harmonic mixture of different audio components.

\begin{figure}[t]
  \centering
 \includegraphics[width=1.0\columnwidth]{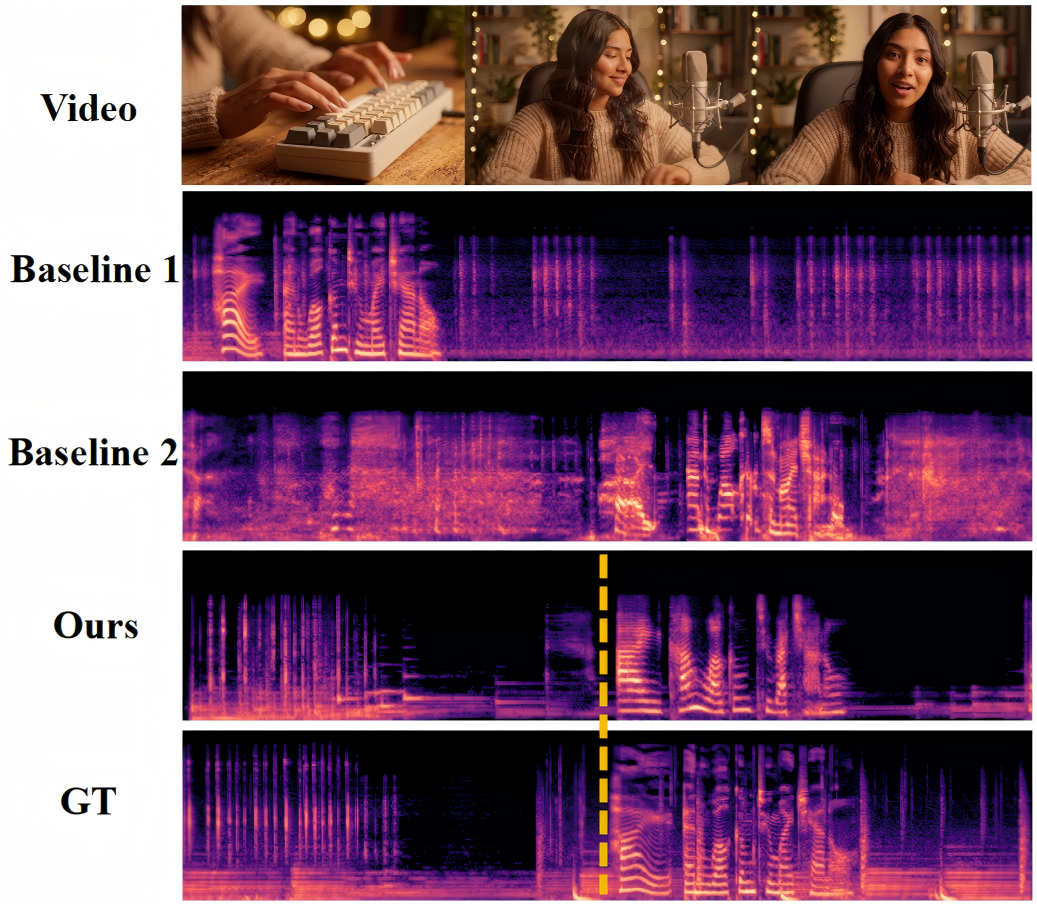}
  \vspace{-1.5em}
  \caption{Qualitative example of mixed soundtrack generation on a Veo3 video. Baseline 1 and Baseline 2 combine MMAudio and AudioX with CosyVoice 3 and LipVoicer, respectively. GT refers to the native synchronized Veo3 audio. The yellow dashed line indicates a key semantic transition point.}
  \label{fig:Foley-Omni-mixed-example}
  \vspace{-0.5em}
\end{figure}

Since comparable open-source systems for complete video soundtrack generation are limited, we construct two compositional baselines from strong single-component generation models.
Specifically, both pipelines employ MMAudio~\cite{cheng2025mmaudio} to generate sound effects from video and caption, and use AudioX~\cite{tian2025audiox} to generate music from caption.
For speech synthesis, we use either CosyVoice 3~\cite{du2025cosyvoice} with the transcript or LipVoicer~\cite{yemini2024lipvoicer} with the video input, leading to two compositional pipelines.
The generated components are then mixed into a soundtrack, providing a comparison between compositional pipelines and the end-to-end generation of Foley-Omni.

For objective evaluation, we leverage CLAP~\cite{wu2023large} and Word Error Rate (WER) to assess adherence across audio, speech, and prompt.
Concurrently, DeSync~\cite{iashin2024synchformer} and the ImageBind (IB) score~\cite{girdhar2023imagebind} are employed to quantify temporal synchronization and audiovisual consistency.
We also conduct human evaluation using Mean Opinion Scores (MOS) to assess audio quality (A-MOS), temporal synchronization (T-MOS), and semantic consistency (S-MOS).
The subjective evaluation set contains 30 samples, including several Veo3-generated videos to probe generalization beyond V2ST-Bench.
Detailed baselines, metrics, and human evaluation protocols are provided in Appendix~\ref{app:evaluation-details}.

\section{Results}

\begin{table*}[ht]
  \centering
  \caption{V2ST-Bench results for complete video soundtrack generation. Ground truth is included as a reference. MOS reliability analysis and confidence intervals are reported in Appendix~\ref{app:main-evaluation-details}.}
  \label{tab:v2s-main}
  \footnotesize
  \setlength{\tabcolsep}{5.2pt}
  \renewcommand{\arraystretch}{1.12}
  \resizebox{\textwidth}{!}{%
  \begin{tabular}{@{}lccccccc@{}}
    \toprule
    Method & CLAP $\uparrow$ & IB $\uparrow$ & WER $\downarrow$ & DeSync $\downarrow$ & A-MOS $\uparrow$ & S-MOS $\uparrow$ & T-MOS $\uparrow$ \\
    \midrule
    GT 
      & 0.30 & 0.36 & 8.03 & 0.14 & 4.33 & 4.37 & 4.42 \\

    MMAudio + CosyVoice 3 + AudioX
      & \underline{0.26} & 0.25 & \underline{10.57} & 0.85 & \underline{2.99} & \underline{3.01} & 2.37 \\
    MMAudio + LipVoicer + AudioX
      & 0.22 & 0.16 & 37.84 & \underline{0.26} & 2.09 & 2.31 & \underline{2.51} \\

    \textbf{Foley-Omni}
      & \textbf{0.27} & \textbf{0.26} & \textbf{7.59} & \textbf{0.16} & \textbf{3.92} & \textbf{4.13} & \textbf{4.14} \\
    \bottomrule
  \end{tabular}%
  }
\end{table*}

Experimental results are presented from three aspects.
First, we evaluate Foley-Omni on complete video soundtrack generation, the core advantage over existing video-based foley generation systems.
Second, we assess its task-level generation abilities on standard benchmarks.
Finally, we conduct ablation studies to analyze the contribution of curriculum learning and synchronization injection.

\subsection{Complete Video Soundtrack Generation}

As reported in Table~\ref{tab:v2s-main}, Foley-Omni consistently outperforms compositional baselines across all metrics on V2ST-Bench. On objective metrics, Foley-Omni achieves the lowest WER and DeSync scores, underscoring its capability to preserve speech intelligibility while maintaining precise temporal alignment with the visual stream. Furthermore, it yields the highest IB and CLAP scores, validating superior audiovisual consistency and semantic relevance. Subjective human evaluations further validate the robust advantages of this unified paradigm. Foley-Omni delivers substantial gains in perceived audio quality (A-MOS), semantic consistency (S-MOS), and temporal synchronization (T-MOS) over compositional baselines.

Specifically, the CosyVoice-based baseline struggles with temporal alignment due to the absence of visual conditioning. In addition, its elevated WER and degraded A-MOS reveal that directly mixing independent tracks introduces cross-component interference, where sound effects and music may suppress speech clarity. 
While the LipVoicer-based baseline yields better synchronization performance, its high WER indicates that video-only speech generation struggles to produce intelligible and linguistically accurate speech. Fundamentally, their degraded T-MOS exposes a core limitation of post-mixing pipelines: independent generation processes struggle to cohesively align diverse audio components with the visual stream.

Figure~\ref{fig:Foley-Omni-mixed-example} provides a qualitative example. 
As highlighted by the yellow dashed line, our model achieves a crisp temporal boundary between the short impulsive patterns of keyboard interaction and the continuous harmonic structures of human speech, precisely mirroring the ground truth. 
In contrast, the spectrograms of both compositional baselines reveal severe temporal smearing and cross-component interference across this transition.  

\begin{table}[t]
  \centering
  \caption{Text-conditioned generation results across TTA, TTS, and TTM tasks. ``--'' indicates unsupported or unreported results.}
  \label{tab:text-main}
  \footnotesize
  \setlength{\tabcolsep}{4.2pt}
  \renewcommand{\arraystretch}{1.10}
  \resizebox{\columnwidth}{!}{%
  \begin{tabular}{@{}llccc@{}}
    \toprule
    Type & Model 
    & CLAP$_{\mathrm{TTA}}$ $\uparrow$ 
    & WER$_{\mathrm{TTS}}$ $\downarrow$ 
    & CLAP$_{\mathrm{TTM}}$ $\uparrow$ \\
    \midrule
    TTA & AudioLDM~2~\cite{liu2024audioldm2}
        & 0.43 & -- & -- \\
        & MMAudio~\cite{cheng2025mmaudio}
        & \textbf{0.49} & -- & -- \\
    \midrule
    TTS & MaskGCT~\cite{wang2025maskgct}
        & -- & 3.03 & -- \\
        & CosyVoice 3~\cite{du2025cosyvoice}
        & -- & \textbf{1.96} & -- \\
    \midrule
    TTM & MusicGen~\cite{copet2023simple}
        & -- & -- & 0.245 \\
    \midrule
    Unified & AudioX~\cite{tian2025audiox}
        & 0.44 & -- & \textbf{0.386} \\
            & UniFlow-Audio~\cite{xu2025uniflow}
        & \underline{0.46} & \underline{2.19} & 0.241 \\
            & \textbf{Foley-Omni}
        & \underline{0.46} & 2.31 & \underline{0.374} \\
    \bottomrule
  \end{tabular}%
  }
\end{table}

\begin{table*}[t]
  \centering
  \caption{Video-to-audio results on VGGSound. All baselines are conditioned on both video and text prompt.}
  \label{tab:V2A-main}
  \footnotesize
  \setlength{\tabcolsep}{5.2pt}
  \renewcommand{\arraystretch}{1.12}
  \resizebox{\textwidth}{!}{%
  \begin{tabular}{@{}lccccccc@{}}
    \toprule
    Model & 
    $\mathrm{FD}_{\mathrm{VGG}}\downarrow$ &
  $\mathrm{FD}_{\mathrm{PASST}}\downarrow$ &
  $\mathrm{KL}_{\mathrm{PANN}}\downarrow$ &
  CLAP $\uparrow$ &
  IS $\uparrow$ &
  IB $\uparrow$ &
  DeSync $\downarrow$ \\
    \midrule
    VTA-LDM~\cite{xu2024video} & 18.77 & 827.57 & 3.45 & -0.04 & 2.01 & 0.06 & 1.17 \\
    FoleyCrafter~\cite{zhang2026foleycrafter} & 2.54 & 137.52 & 2.32 & 0.19 & 15.02 & 0.25 & 1.23 \\
    MMAudio~\cite{cheng2025mmaudio} & \textbf{1.27} & 102.93 & \underline{1.99} & \underline{0.21} & \textbf{15.76} & \underline{0.31} & 0.57 \\
    HunyuanVideo-Foley~\cite{shan2025hunyuanvideo} & 2.18 & \textbf{79.07} & 2.02 & \textbf{0.23} & \underline{15.49} & \textbf{0.32} & \underline{0.55} \\
    \midrule
    \textbf{Foley-Omni} & \underline{1.57} & \underline{101.40} & \textbf{1.92} & \underline{0.21} & 14.00 & 0.28 & \textbf{0.50} \\
    \bottomrule
  \end{tabular}%
  }
\end{table*}

\begin{table*}[t]
  \centering
  \caption{VisualTTS Setting 1: GRID seen-speaker results. All baselines are conditioned on both video and transcript.}
  \label{tab:visualtts-grid}
  \footnotesize
  \setlength{\tabcolsep}{4pt}
  \renewcommand{\arraystretch}{1.12}
  \resizebox{\textwidth}{!}{
  \begin{tabular}{l c c c c c c c c}
    \toprule
    Method & WER $\downarrow$ & Spk. Sim. $\uparrow$ & UTMOS $\uparrow$ & MCD $\downarrow$ & MCD-D $\downarrow$ & MCD-DS $\downarrow$ & LSE-C $\uparrow$ & LSE-D $\downarrow$ \\
    \midrule
    GT & 18.0 & 100.0 & 3.96 & 0.00 & 0.00 & 0.00 & 7.25 & 6.71 \\
    HPMDubbing~\cite{cong2023learning} & 44.1 & 72.0 & 2.13 & 12.45 & 8.11 & 8.23 & \underline{5.87} & 9.11 \\
    ProDubber~\cite{zhang2025prosody} & \textbf{12.1} & 79.7 & \underline{3.74} & 11.74 & 6.52 & \underline{6.24} & 2.97 & 10.21 \\
    EmoDubber~\cite{cong2025emodubber} & 16.9 & \underline{83.9} & \textbf{3.98} & \textbf{9.03} & \textbf{5.89} & \textbf{5.97} & \textbf{7.14} & \textbf{6.75} \\
    \textbf{Foley-Omni} & \underline{15.3} & \textbf{84.1} & 3.07 & \underline{10.41} & \underline{6.28} & 6.35 & 5.24 & \underline{8.67} \\
    \bottomrule
  \end{tabular}
  }
\end{table*}

\begin{table}[t]
  \centering
  \caption{VisualTTS Setting 2: LRS2 zero-shot results.}
  \label{tab:visualtts-lrs2}
  \footnotesize
  \setlength{\tabcolsep}{3.4pt}
  \renewcommand{\arraystretch}{1.08}
  \resizebox{\columnwidth}{!}{%
  \begin{tabular}{@{}llcccc@{}}
    \toprule
    Type & Method & WER $\downarrow$ & Spk. Sim. $\uparrow$ & UTMOS $\uparrow$ & MCD-DS $\downarrow$ \\
    \midrule
    GT & GT & 8.93 & 100.0 & 3.14 & 7.59 \\
    \midrule
    \multirow{2}{*}{Video}
      & LipVoicer & \underline{22.5} & 62.1 & 2.16 & 9.52 \\
      & Faces2Voices & 31.5 & \textbf{65.0} & \textbf{3.88} & \underline{8.26} \\
    \midrule
    \multirow{2}{*}{\shortstack[c]{Text\\+Video}}
      & EmoDubber & 110.0 & 10.3 & 1.70 & 8.68 \\
      & \textbf{Foley-Omni} & \textbf{13.0} & \underline{64.6} & \underline{2.67} & \textbf{8.24} \\
    \bottomrule
  \end{tabular}%
  }
\end{table}

\subsection{Task-Level Synthesis}

\paragraph{Text-conditioned generation.}
Table~\ref{tab:text-main} reports text-conditioned generation results on TTA, TTS, and TTM tasks.
Foley-Omni achieves competitive performance compared to task-specific expert systems and unified audio generation systems.
These results indicate that Foley-Omni preserves broad text-conditioned generation capabilities across speech, sound effects, and music, which serve as the foundation for generating complete video soundtrack.

\paragraph{V2A}
Results on the standard V2A benchmark VGGSound are shown in Table~\ref{tab:V2A-main}.
Foley-Omni achieves the best DeSync score on VGGSound, demonstrating strong temporal alignment between generated sounds and visual events.
This improvement may come from joint training with VisualTTS data, which provides better synchronization for speaking scenes in the VGGSound test set.
In terms of other metrics, Foley-Omni also remains highly competitive, showing strong audio fidelity and semantic consistency with the input video.

\paragraph{VisualTTS}
For VisualTTS, Tables~\ref{tab:visualtts-grid} and~\ref{tab:visualtts-lrs2} report results on GRID~\cite{cooke2006grid} and LRS2~\cite{afouras2018lrs2}.
The GRID test speakers are seen during training, while the LRS2 test speakers are unseen, corresponding to a zero-shot evaluation setting.
On GRID, Foley-Omni achieves the best speaker similarity and the second-best WER, showing strong speaker preservation and speech intelligibility.
Although expert VisualTTS systems like EmoDubber obtain better results on LSE-based~\cite{chung2017out} synchronization metrics, they are trained only on GRID and rely on complex lip-region preprocessing.
Their high WER in the LRS2 zero-shot setting suggests limited generalization to unseen speakers and more realistic videos.
To further examine performance under realistic scenarios, we also compare with video-only speech generation models, LipVoicer and Faces2Voices~\cite{kim2025faces}.
LipVoicer shows strong zero-shot robustness and is therefore used in our compositional pipelines, but these video-only methods inherently suffer from elevated WER due to the absence of textual condition.
In contrast, Foley-Omni achieves the best WER with competitive speaker similarity, demonstrating a strong balance between intelligibility, speaker preservation, and generalization.

\subsection{Ablation Study}

We evaluate the necessity of our core design choices through two independent ablation settings: (1) eliminating the synchronization feature additive path, and (2) replacing the curriculum learning strategy with single-stage joint optimization of all tasks under the same training steps. 

As reported in Table~\ref{tab:ablation}, removing the additive synchronization path primarily compromises audiovisual alignment, causing noticeable regressions in IB. 
This underscores the contribution of additive synchronization feature fusion in capturing visual timing for precise alignment. 
Furthermore, single-stage joint training results in a substantial deterioration of WER across both GRID and V2ST-Bench. 
This validates that multi-task optimization from scratch induces destructive cross-task conflict, severely degrading the semantic clarity of the generated speech.
These findings highlight the importance of both design choices in our method.

\begin{table}[t]
  \centering
  \caption{Ablation results. "Single-stage training" eliminates the curriculum learning strategy; ``w/o $\mathbf{Z}_{\mathrm{sync}}$ '' removes the additive synchronization path.}
  \label{tab:ablation}
  \footnotesize
  \setlength{\tabcolsep}{5pt}
  \renewcommand{\arraystretch}{1.12}
  \resizebox{\columnwidth}{!}{%
  \begin{tabular}{lcccc}
    \toprule
    Variant & FD$_{\text{VGG}}$$\downarrow$  & WER$_{\text{GRID}}$ $\downarrow$ & IB$_{\text{V2ST}}$ $\uparrow$ & WER$_{\text{V2ST}}$ $\downarrow$ \\
    \midrule
    Single-stage training & \underline{1.73} & 27.4 & \underline{0.24} & 29.29 \\
    w/o $\mathbf{Z}_{\mathrm{sync}}$  & 2.21 & \underline{18.9} & 0.22 & \underline{12.40} \\
    Full model & \textbf{1.57} & \textbf{15.3} & \textbf{0.26} & \textbf{7.59} \\
    \bottomrule
  \end{tabular}}
\end{table}

\section{Conclusion}

We present Foley-Omni, a unified multimodal generation model that extends audio generation from isolated tasks to complete video soundtrack synthesis.
Foley-Omni combines structured text conditions, semantic and synchronization-aware video features to jointly generate speech, sound effects, and music.
We also introduce an audiovisual data curation pipeline and V2ST-Bench to support reproducible research on realistic mixed soundtrack generation.
Experiments show that Foley-Omni remains competitive on task-level synthesis while achieving clear gains on full soundtrack generation, especially in intelligibility, audiovisual consistency, and perceived quality.


\section*{Limitations}

Our current study still has several limitations that point to promising directions for future research. First, the model currently generates a single final mixture rather than explicitly exposing fine-grained control interfaces for source balance, speaker prominence, or music intensity, necessitating the exploration of more controllable soundtrack editing mechanisms. 
Second, existing unified systems that jointly support visual speech generation and V2A are not open-sourced. As a result, our main comparison relies on reproducible compositional pipelines built from strong task-level models.
We hope that the release of Foley-Omni and V2ST-Bench will facilitate future research and enable more systematic comparisons in this emerging direction.
Finally, the perceptual clarity of the generated speech is occasionally constrained by the diverse multi-speaker data used during training.
We plan to introduce reference audio conditions in future iterations to reduce multi-speaker interference and improve speech generation quality.

\section*{Ethics Statement}

Our research involves human subjective evaluation to assess the quality of the generated audiovisual content. We ensured that all human evaluation protocols strictly adhered to ethical guidelines. All participants were informed about the purpose of the study and provided explicit consent prior to participation. Furthermore, annotators were fairly compensated above the local minimum wage for their time and effort. The entire evaluation process was conducted anonymously, and no Personally Identifiable Information or sensitive data was collected, stored, or distributed. Internal data used in training are collected under appropriate usage agreements.  We acknowledge the potential misuse of our models for creating deepfakes, and strongly advocate for responsible use alongside the development of detection tools.

\bibliography{custom}

\clearpage   
\appendix
\section*{Appendix}

\section{Details of the Audiovisual Data Curation Pipeline}
\label{app:pipeline-details}

This appendix provides the comprehensive technical specifications, hyper-parameters, and operational prompts utilized in our audiovisual data curation pipeline (Sec.~\ref{sec:datapipeline}).

\subsection{Multimodal Data Filtering Metrics}

To construct a high-quality pre-training corpus, we aggregate large-scale audiovisual data from both raw internet video collections and established open-source datasets. Because these corpora inherently contain noisy, heavily compressed, or weakly aligned samples, we implement a rigorous automated filtering pipeline prior to the dense annotation stage. 

Each candidate clip must satisfy a joint set of constraints across visual quality, acoustic fidelity, and cross-modal alignment. The specific metrics and their corresponding operational thresholds are detailed in Table~\ref{tab:filtering-thresholds}. Only video-audio pairs that strictly pass all criteria are preserved for subsequent dense annotation.

\begin{table}[htbp]
  \centering
  \caption{Filtering metrics and operational thresholds used in our audiovisual data curation pipeline.}
  \label{tab:filtering-thresholds}
  \footnotesize
  \setlength{\tabcolsep}{5pt}
  \renewcommand{\arraystretch}{1.12}
  \begin{tabularx}{\columnwidth}{@{}lXc@{}}
    \toprule
    \textbf{Dimension} & \textbf{Metric} & \textbf{Threshold} \\
    \midrule
    \multirow{3}{*}{\textbf{Visual}}
      & Resolution & $\geq$ 480p \\
      & Bitrate & $\geq$ 1 Mbps \\
      & Motion score & $[0.1, 3.2]$ \\
    \midrule
    \textbf{Audio}
      & Audio Quality ~\cite{tjandra2025meta} & $\geq 0.6$ \\
    \midrule
    \multirow{2}{*}{\textbf{Alignment}}
      & IB score ~\cite{girdhar2023imagebind}& $\geq 0.3$ \\
      & Sync score ~\cite{iashin2024synchformer}& $\geq 0.2$ \\
    \bottomrule
  \end{tabularx}
\end{table}

\subsection{Detection and Annotation}
For clips that pass the filtering stage, we deploy Gemini 2.5 Pro~\cite{comanici2025gemini} to perform multimodal joint detection and structured annotation simultaneously. The model is fed with both the video frames and the corresponding audio track. It is explicitly instructed to first detect whether a specific audio component (\textit{Speech}, \textit{Sound Effects}, or \textit{Music}) is physically present in the clip, and if so, generate the corresponding descriptive caption.

The systematic prompt template used for this joint detection and annotation task is detailed in Table~\ref{box:gemini-prompt}.

\subsection{Acoustic Post-Verification}
To guarantee the reliability of the generated annotations, we implement an automated acoustic post-verification step. We utilize the Bandit~\cite{watcharasupat2024remastering} model to separate the raw audio waveform into three distinct stems: speech ($a_{\text{words}}$), sound effects ($a_{\text{audio}}$), and background music ($a_{\text{music}}$). 

For each separated stem $a_c$, where $c \in \{\text{words}, \text{audio}, \text{music}\}$, we compute its average Root-Mean-Square (RMS) energy in decibels:
\begin{equation}
E(a_c) = 10 \log_{10} \left( \frac{1}{N} \sum_{n=1}^{N} a_c[n]^2 \right),
\end{equation}
where $N$ denotes the total number of audio samples in the clip.

An annotation predicted by the multimodal model is retained if and only if the corresponding audio stem exceeds a predefined energy threshold:
\begin{equation}
E(a_c) > -35 \text{ dB}.
\end{equation}

If the energy $E(a_c)$ falls below $-35\text{ dB}$, we assume the respective audio component is physically absent or negligible. Consequently, the model's textual prediction for that specific category is discarded. We choose -35 dB as a conservative threshold based on manual inspection of a small validation subset, where stems below this level are typically inaudible or negligible in the mixture.  This hard-gating mechanism effectively mitigates visual hallucination and ensures absolute supervision fidelity.

\section{Details of V2ST-Bench}
\label{app:v2st-details}

\begin{figure*}[t]
  \centering
  \includegraphics[width=\textwidth]{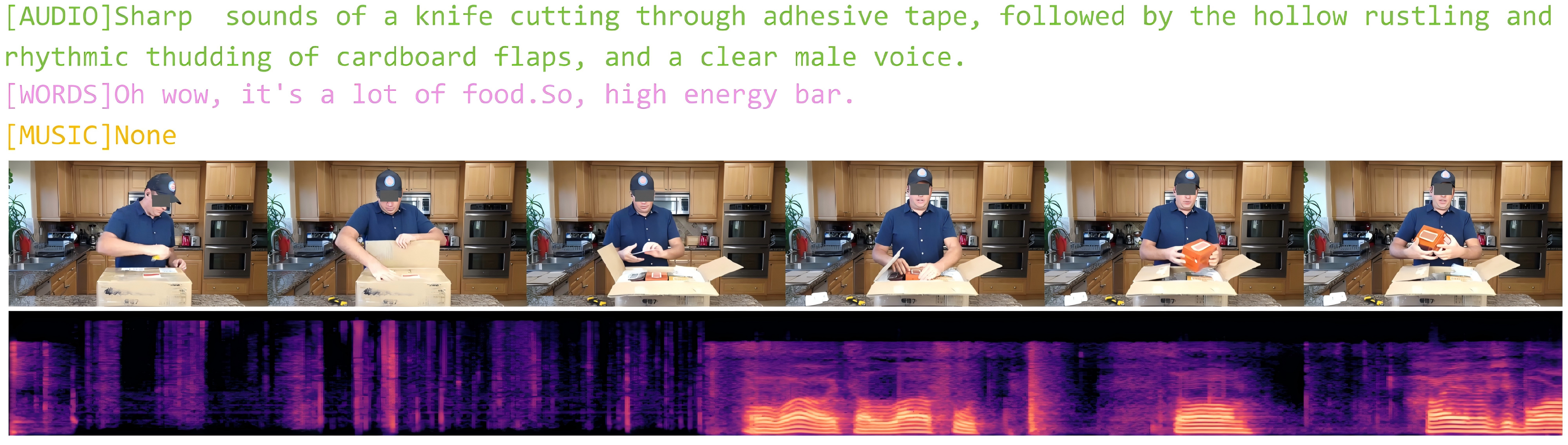}
  \caption{An illustrative example from V2ST-Bench. The figure provides a representative sample, illustrating the structured textual prompts  alongside the visual sequence with facial privacy preserved  and the corresponding acoustic representation.}
  \label{fig:v2st-example}
\end{figure*}

V2ST-Bench consists of video clips ranging from 5 to 10 seconds, each paired with independent, structured text annotations for the spoken transcript, sound effects, and background music. Table~\ref{tab:v2st-bench-stats} details the exact quantitative distribution of these overlapping audio combinations. Figure~\ref{fig:v2st-example} provides a representative annotated sample, illustrating the structured textual prompts alongside the visual sequence and acoustic representation. We will release the structured annotations, metadata, and processing scripts; for web videos whose redistribution is not permitted, we will provide URLs and metadata instead of raw video files.

\begin{table}[htbp]
\centering
\caption{Detailed composition of audio combinations in V2ST-Bench.}
\label{tab:v2st-bench-stats}

\footnotesize 
\setlength{\tabcolsep}{8pt}
\begin{tabular}{lc}
\toprule
\textbf{Audio Combination} & \textbf{Count} \\
\midrule
Speech + Sound Effects & 150 \\
Speech + Music & 120 \\
Speech + Sound Effects + Music & 30 \\
\midrule
\textbf{Total} & \textbf{300} \\
\bottomrule
\end{tabular}
\end{table}

\section{Training}
\begin{table*}[t]

  \centering

  \caption{Training data sources and task grouping. Audiovisual data are processed by the curation pipeline described in Section~\ref{sec:datapipeline}.}
  \label{tab:data-source}
  \footnotesize
  \setlength{\tabcolsep}{6pt}
  \renewcommand{\arraystretch}{1.12}
  \begin{tabular}{@{}lllr@{}}
    \toprule
    Data group & Task type & Dataset & Hours \\
    \midrule
    \multirow{3}{*}{Text--audio}
      & TTS & LJSpeech, LibriTTS, and internal speech data & 1253 \\
      & TTA & AudioCaps, Freesound& 912 \\
      & TTM & MusicCaps, MusicBench, AudioSet music subset & 139 \\
    \midrule
    \multirow{3}{*}{Video--text--audio}
      & VisualTTS & Chem, GRID, LRS2, SpeakerVid, Talkvid & 1980 \\
      & V2A & VGGSound, Kling-Foley, and internal video & 403 \\
      & V2ST & Internal data  and SpeakerVid & 216 \\
    \bottomrule

  \end{tabular}

\end{table*}

\subsection{Training Data}
\label{app:data-details}

Our model is trained on a diverse mixture of publicly available datasets and internal collections, as detailed in Table~\ref{tab:data-source}. The specific data composition for each task type is formulated as follows.

\paragraph{TTS} This domain includes approximately 1.3k hours of text-conditioned speech data, leveraged from LJSpeech~\cite{ljspeech17}, LibriTTS~\cite{zen2019libritts}, and our internal speech repository.

\paragraph{TTA} This subset contains approximately 0.9k hours of audio paired with textual descriptions, primarily sourced from AudioCaps~\cite{kim2019audiocaps} and Freesound~\cite{mei2024wavcaps}.

\paragraph{TTM} This component comprises approximately 0.1k hours of music tracks, combining data from MusicCaps~\cite{agostinelli2023musiclm}, MusicBench~\cite{melechovsky2024mustango}, the music subset of AudioSet~\cite{gemmeke2017audio}, and internal music assets.

\paragraph{V2A.}
This group contains approximately 0.4k hours of video-audio data for video-conditioned sound effect generation. The data are curated from VGGSound~\cite{chen2020vggsound}, Kling-Foley~\cite{wang2025kling}, and internal audiovisual corpora. Except for the held-out test sets, all samples are processed by the pipeline described in Section~\ref{sec:datapipeline}.

\paragraph{VisualTTS.}
This group contains approximately 1.9k hours of lip-synchronized or video-conditioned speech data. The data are collected from Chem~\cite{chen2022v2c}, GRID~\cite{cooke2006grid}, LRS2~\cite{afouras2018lrs2}, SpeakerVid~\cite{zhang2025speakervid}, Talkvid~\cite{chen2025talkvid}, and internal speaker corpora. Similarly, all samples are filtered and standardized using the  pipeline in Section~\ref{sec:datapipeline}.

\paragraph{V2ST.}
This group contains approximately 0.2k hours of mixed audiovisual data for complete soundtrack generation. The data mainly come from SpeakerVid and internal audiovisual corpora after curation, and each retained sample contains at least two of the following components: speech, sound effects, and music. This group is used to improve mixed-source generation, where multiple audio components need to coexist within the same video.

\subsection{Training Configurations}
\label{app:trainconfig}

All experiments are conducted on 8 NVIDIA H200 GPUs with a global batch size of 32. We use AdamW as the optimizer. The base learning rate is set to $5 \times 10^{-5}$ for the first two stages, and reduced to $2 \times 10^{-5}$ for mixed-source finetuning.

The model is first trained for 5 epochs on approximately 0.7M text-audio pairs covering TTA, TTS, and TTM. It is then trained for 3 epochs on video-text-audio pairs covering V2A and VisualTTS. The final stage consists of 2 epochs of finetuning on curated V2ST samples with coexisting audio components. Crucially, to mitigate catastrophic forgetting and preserve proficiency in individual tasks, we employ a data replay strategy during this final stage by integrating 100 hours of data from each prior single-task domain. The full training process contains approximately 50k optimization steps, as summarized in Table~\ref{tab:training-config}.

\begin{table}[t]
  \centering
  \caption{Training configuration for the progressive training strategy.}
  \label{tab:training-config}
  \footnotesize
  \setlength{\tabcolsep}{5pt}
  \renewcommand{\arraystretch}{1.12}
  \resizebox{\columnwidth}{!}{%
  \begin{tabular}{@{}lcc@{}}
    \toprule
    Stage & Task groups & Epochs \\
    \midrule
    Text-audio pretraining
      & TTA, TTS, TTM & 5 \\
    Video-conditioned expansion
      & V2A, VisualTTS & 3 \\
    Complete soundtrack finetuning
      & V2ST & 2 \\
    \bottomrule
  \end{tabular}%
  }
\end{table}

\section{Evaluation Setup}
\label{app:evaluation-details}

\subsection{Details of V2ST Evaluation Protocols}
\label{app:main-evaluation-details}

\paragraph{Baseline Implementations.}
To ensure a fair comparison, the textual conditions fed into the compositional baselines are strictly aligned with the structured annotations of V2ST-Bench. 
Specifically, MMAudio is conditioned on the video and sound effect captions, while AudioX is driven by the music prompts. For speech synthesis, CosyVoice 3 utilizes the spoken transcript alongside a reference speech sample, whereas LipVoicer relies solely on the visual track. During the post-hoc mixing phase, we observed that MMAudio occasionally hallucinates unintended human vocals. To prevent these artifacts from interfering with the dedicated speech track, we apply a vocal-removal preprocessing step to the MMAudio outputs before fusing the generated components into the final soundtrack.

\begin{figure*}[t]
  \centering
  \includegraphics[width=0.95\textwidth]{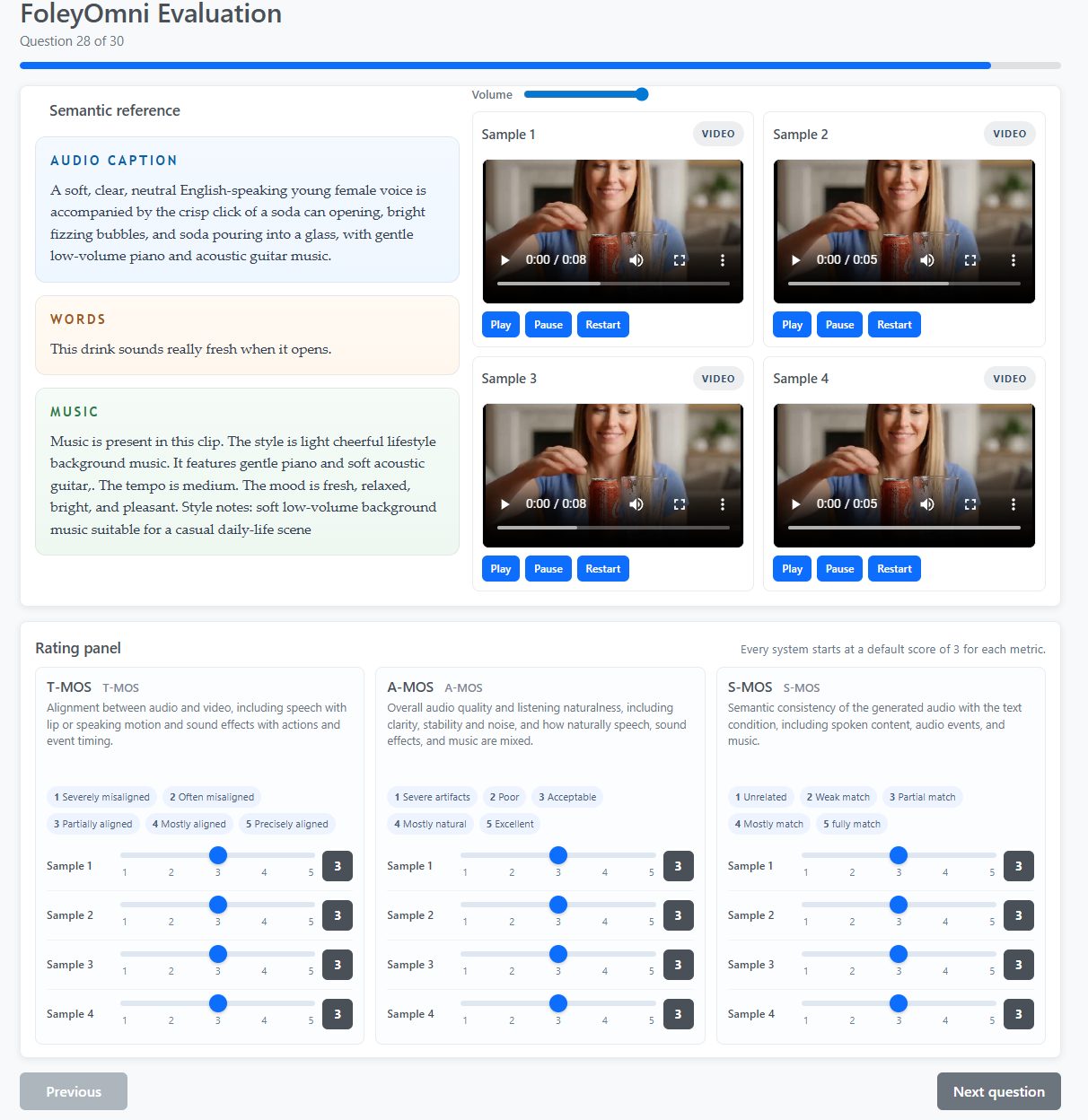}
  \caption{A representative screenshot of the subjective evaluation interface used for the Foley-Omni MOS test.}
  \label{fig:mos-interface}
\end{figure*}

\paragraph{Subjective Evaluation Protocol.}
We conducted a Mean Opinion Score (MOS) test comprising 30 video samples: 20 representative clips selected from V2ST-Bench and 10 Veo3-generated videos to probe generalization. 
Twenty evaluators with university-level listening proficiency participated in the assessment within a quiet environment, rating each sample on a standard 1--5 scale. 
Participants evaluated A-MOS for overall audio quality, T-MOS for temporal synchronization between audio events and visual cues, and S-MOS for semantic adherence to the provided text conditions. 
For each sample and each system, the generated audios were presented in a randomized order without revealing system identities.
A screenshot of the subjective evaluation interface used in our experiments is provided in Figure~\ref{fig:mos-interface}.

Each system-metric pair contains 600 valid ratings.
We compute 95\% confidence intervals for MOS scores using the $t$ distribution over valid ratings; bootstrap percentile intervals give nearly identical results.
To assess inter-rater reliability, we compute ICC(2,1) for individual evaluator agreement and ICC(2,k) for the reliability of evaluator-averaged scores, following Koo and Li~\cite{koo2016guideline}.
Across T-MOS, A-MOS, and S-MOS, ICC(2,1) ranges from 0.573 to 0.582, while ICC(2,k) ranges from 0.965 to 0.969.
Cronbach's $\alpha$ ranges from 0.832 to 0.840, indicating good internal consistency.
The detailed MOS confidence intervals are reported in Table~\ref{tab:mos-ci}.

\begin{table}[t]
  \centering
  \caption{MOS results reported as mean $\pm$ 95\% confidence interval.}
  \label{tab:mos-ci}
  \footnotesize
  \setlength{\tabcolsep}{4.5pt}
  \renewcommand{\arraystretch}{1.12}
  \resizebox{\columnwidth}{!}{%
  \begin{tabular}{lccc}
    \toprule
    Method & A-MOS & S-MOS & T-MOS \\
    \midrule
    GT 
      & 4.33 $\pm$ 0.07 
      & 4.37 $\pm$ 0.07 
      & 4.42 $\pm$ 0.06 \\
    MMAudio + CosyVoice 3 + AudioX 
      & 2.99 $\pm$ 0.08 
      & 3.01 $\pm$ 0.09 
      & 2.37 $\pm$ 0.09 \\
    MMAudio + LipVoicer + AudioX 
      & 2.09 $\pm$ 0.08 
      & 2.31 $\pm$ 0.09 
      & 2.51 $\pm$ 0.10 \\
    Foley-Omni 
      & 3.92 $\pm$ 0.07 
      & 4.13 $\pm$ 0.07 
      & 4.14 $\pm$ 0.07 \\
    \bottomrule
  \end{tabular}%
  }
\end{table}

\subsection{Details of Task-Level Evaluation}
We provide detailed evaluation settings for the task-level benchmarks used in the main paper. The evaluation covers text-conditioned generation, video-to-audio generation, and visual speech synthesis. For each task, we follow commonly used benchmarks, baselines, and metrics from prior work.

\paragraph{TTA}
For TTA, we evaluate on the AudioCaps test set~\citep{kim2019audiocaps}. This task measures whether a model can generate general audio events from natural language descriptions. We compare Foley-Omni with representative text-to-audio systems, including AudioLDM~2, as well as recent general audio generation models. Following AudioLDM~2, we report CLAP~\cite{wu2023large} score and Fr\'echet Distance (FD). CLAP score measures the semantic similarity between the input text and generated audio in a joint text-audio embedding space. FD measures the distributional distance between generated and reference audio features.

\paragraph{TTS}
For TTS, we evaluate on LibriSpeech-PC~\citep{panayotov2015librispeech}. This task focuses on whether generated speech accurately preserves the linguistic content of the input transcript. We compare Foley-Omni with strong TTS systems, including MaskGCT~\cite{wang2025maskgct} and CosyVoice. We use word error rate (WER) by Whisper-large-v3~\citep{radford2023robust} as the primary metric.  Lower WER indicates higher speech intelligibility and better faithfulness to the input text.

\paragraph{TTM}
For TTM, we evaluate on MusicCaps~\citep{agostinelli2023musiclm}. We compare Foley-Omni with representative music generation system MusicGen~\cite{copet2023simple} and recent unified audio generation models. We also report CLAP score to measure text-music semantic consistency and FD to evaluate the distributional quality of generated music. These metrics are used to assess whether the unified model preserves music generation ability while also supporting speech and sound-effect generation.

\paragraph{V2A}
For V2A, we evaluate on VGGSound~\citep{chen2020vggsound}. This task requires the model to generate sound effects that are semantically related to the video and temporally aligned with visual events. We compare Foley-Omni with representative V2A systems, including VTA-LDM, FoleyCrafter, MMAudio, and HunyuanVideo-Foley. FD and KL divergence assess the distributional quality of generated audio. CLAP measures semantic consistency between generated audio and text or category labels. IS reflects the diversity and recognizability of generated sounds. ImageBind score (IB) evaluates video-audio semantic similarity~\citep{girdhar2023imagebind}. DeSync estimates temporal alignment using Synchformer, where lower values indicate better audiovisual alignment. All evaluation settings follow MMAudio.

\paragraph{VisualTTS on GRID.}
For VisualTTS, we first evaluate the seen-speaker setting on GRID~\citep{cooke2006grid}. The GRID dataset contains videos from 33 speakers, with 1,000 utterances per speaker. The test set consists of 3,280 total samples. This setting mainly evaluates whether the model can generate intelligible and speaker-consistent speech when the speaker identity has been seen before. We compare Foley-Omni with representative visual dubbing systems, including HPMDubbing, ProDubber, and EmoDubber. For a fair comparison, we utilize their respective models trained on the GRID dataset, taking both reference audio and text as inputs during testing. We follow the evaluation metrics established in previous works. WER measures speech intelligibility. Speaker similarity is computed as the cosine similarity between speaker embeddings of generated and reference speech. UTMOS~\cite{saeki2022utmos} estimates speech naturalness. MCD-based metrics measure spectral distortion between generated and reference speech under different alignment or silence-removal settings. LSE-C and LSE-D are lip-synchronization metrics derived from SyncNet models, measuring synchronization confidence and feature distance, respectively.

\paragraph{VisualTTS on LRS2.}
We further evaluate zero-shot VisualTTS on LRS2~\citep{afouras2018lrs2}. Compared with GRID, LRS2 contains more realistic talking-face videos from television programs, with larger variations in speaker identity, pose, illumination, background, and recording conditions. We use 800 test samples whose speakers are unseen during training. This setting evaluates zero-shot generalization under more natural video conditions. In addition to text-and-video guided dubbing models, we also include video-only speech generation methods such as LipVoicer and Faces2Voices for a more complete comparison. This is useful because video-only methods can generalize to unseen speakers but do not receive transcript input, which usually leads to higher WER. We report WER, speaker similarity, UTMOS, and MCD-DS, covering speech intelligibility, speaker preservation, naturalness, and spectral distortion.

\begin{table*}[htbp]
\centering
\caption{Streamlined system prompt blueprint for audio-centric soundtrack annotation.}
\label{box:gemini-prompt}
\begin{tabular}{|p{0.95\textwidth}|}
\hline
\rowcolor[gray]{0.95} \textbf{System Prompt for Streamlined Soundtrack Annotation} \\ \hline
\small
\textbf{[Role \& Core Philosophy]} \\
You are an experienced multimedia describer: in one continuous paragraph identify every prominent sound source, its material, and acoustic cue. Avoid subjective interpretations. You must use the video as supplementary evidence to support or clarify audio-based judgments, especially when sound sources or materials are ambiguous. Do not record any purely visual properties (such as clothing colors, facial aesthetics, or lighting) that do not contribute to sound generation. \\\\

\textbf{[Output Structure Rules]} \\
Your output must consist exactly of the following blocks: \\\\
\textbf{General Captions:} \\
Generate ONE continuous narrative paragraph (under 512 tokens). Interleave spoken dialogue using the explicit syntax: \texttt{[WORDS]Verbatim dialogue[END\_WORDS]}. End the paragraph with a concise sound summary enclosed in \texttt{[AUDIO]...[END\_AUDIO]}. \\\\
\textbf{Sound Effect Analysis:} \\
\quad $\bullet$ Detection: Sound Effect detected: [Yes/No] \\
\quad $\bullet$ Type Identification (List): Identify the category of the sound effect (e.g., Footsteps, Door Slam). \\\\
\textbf{Music Analysis:} \\
\quad $\bullet$ Detection: Music detected: [Yes/No] \\
\quad $\bullet$ Music Caption: Describe the background music using 3 to 6 short, natural sentences. Start with "This song features...". Naturally characterize the acoustic elements by detailing Instruments (e.g., steel-string guitar), Genre (e.g., Classical/Pop), Tempo (e.g., Fast/Medium/Slow), Emotional Tone (e.g., Happy/Melancholic), and Style Notes (e.g., blues improvisation). Explicitly state if no voices are audible. (e.g., This song features a steel-string guitar playing a gentle pop rhythm. The tempo is slow, creating a calm and slightly melancholic emotional tone with subtle blues improvisations. There are no voices in this song. ) \\\\

\textbf{[Prohibited Errors]} \\
$\bullet$ Avoid repetition and redundancy. Merge descriptions for recurring or cyclical content. \\
$\bullet$ Simplify technical language. Avoid specialized terms (e.g., no "Hz", "dB", or abstract descriptors.) \\
$\bullet$ Skip unnecessary details (e.g., faint background noise) unless they become prominent. \\\\

\textbf{[Operational Example]} \\
\textbf{General Captions:} \\
A female speaker delivers dialogue distinctly in an indoor workspace, \texttt{[WORDS]}Usually, my health advice gets cut off by the meeting bell.\texttt{[END\_WORDS]} She continues seamlessly, adding, \texttt{[WORDS]}But this ten-second clip is just enough time to remind busy women that one pack covers all your daily nutrient needs.\texttt{[END\_WORDS]} Throughout the entire clip, a gentle, rhythmic sound of typing on a computer keyboard. \texttt{[AUDIO]} A soft office hum blends with gentle, rhythmic typing on keyboards, accompanied by a clear female voice. \texttt{[END\_AUDIO]} \\\\
\textbf{Sound Effect Analysis:} \\
\quad Detection: Yes \\
\quad Type Identification: [ "gentle rhythmic typing"] \\
\textbf{Music Analysis:} \\
\quad Detection: Music detected: No \\
\hline
\end{tabular}
\end{table*}

\end{document}